\documentclass[english]{article}
\usepackage{graphicx}
\usepackage{babel}
\begin{document}

\title{Dynamic critical exponents of Swendsen-Wang and Wolff
algorithms by nonequilibrium relaxation}
\author{Jianqing Du$^1$, Bo Zheng$^1$, and Jian-Sheng Wang$^2$\\
 \\
$^1$Zhejiang Institute of Modern Physics, Zhejiang University,\\
Hangzhou 310027, P. R. China\\
$^2$Department of Physics, National University of Singapore,\\
Singapore 117542, Republic of Singapore}

\date{2 March 2006}

\maketitle
\begin{abstract}
With a nonequilibrium relaxation method, we calculate the dynamic
critical exponent $z$ of the two-dimensional Ising model for the
Swendsen-Wang and Wolff algorithms.  We examine dynamic relaxation
processes following a quench from a disordered or an ordered initial
state to the critical temperature $T_c$, and measure the exponential
relaxation time of the system energy. For the Swendsen-Wang algorithm
with an ordered or a disordered initial state, and for the Wolff
algorithm with an ordered initial state, the exponential relaxation
time fits well to a logarithmic size dependence up to a lattice size
$L=8192$.  For the Wolff algorithm with a disordered initial state, we
obtain an effective dynamic exponent $z_{\rm exp}=1.19(2)$ up to
$L=2048$.  For comparison, we also compute the effective dynamic
exponents through the integrated correlation times. In addition, an
exact result of the Swendsen-Wang dynamic spectrum of a one-dimension
Ising chain is derived.
\end{abstract}

\section{Introduction}

In recent two decades, cluster algorithms have played an important
role in statistical physics due to their reduced critical slowing
down, improved computational efficiency, and interesting dynamical
properties. Among these dynamical properties, the dynamic critical
exponent $z$ is the center of attraction, which describes the
divergent correlation time.

There are various ways to calculate the dynamic critical exponent $z$,
for example, through the exponential decay of the time correlation of
a finite system in equilibrium \cite {william,wansleben}, or from the
dynamic scaling behavior in nonequilibrium states
\cite{li,luo98,zhe98}.  In calculating the time correlation in
equilibrium, the difficulty is that one can hardly reach a very large
lattice.  The advantage for computing the dynamic exponent from a
nonequilibrium relaxation process is that the finite size effect is
more or less negligible, since the spatial correlation length is small
in the early stage of the dynamic relaxation.  Such a nonequilibrium
approach, however, becomes subtle for the cluster algorithms, for the
dynamic exponent $z$ is believed to be close to zero. In addition, it
is also somewhat controversial in defining a Monte Carlo time for the
Wolff algorithm.

In a recent article \cite {gunduc}, an attempt is made to estimate the
dynamic exponent $z$ of the Wolff algorithm from the finite size
scaling behavior in a nonequilibrium state.  A vanishing $z$ value is
claimed. To our understanding, however, the identification of the
dynamic scaling behavior there seems to be not appropriate. On the
other hand, although the cluster algorithms are known to be very
efficient in reducing critical slowing down with a small dynamic
exponent $z$, it has not been rigorously studied what precise values
the dynamic exponent $z$ takes for different variants of the
algorithms. This is important in theory and application of the cluster
algorithms.

In this paper, we will calculate the dynamic exponent $z$ for the
Wolff \cite{wolff} and Swendsen-Wang \cite{sw} algorithms,
respectively, through the exponential relaxation time and integrated
correlation time \cite{landau-binder} of the system energy in
nonequilibrium relaxation processes.  Compared with methods based on
computations of time correlation functions in equilibrium, much larger
system sizes can be reached in the nonequilibrium dynamic approach,
especially in the case of the two-dimensional Ising model, where the
system energy in the equilibrium state is known exactly.  Compared with
the methods in Ref. \cite {gunduc}, the system energy in our
calculations is self-averaged and thus much less fluctuating in
simulations of large lattices.

The paper is organized as follows.  In Sec.~2, the general theory of
the spectrum of the Monte Carlo dynamics is described, and in Sec.~3,
an exact calculation of the spectrum of the one-dimensional Ising
chain is formulated. In Sec.~4, simulation setup is discussed in
detail, and in Sec.~5, numerical results are presented.

\section{Spectrum of Monte Carlo Dynamics}

In order to justify our method, we first look at the spectrum of a
Markov chain Monte Carlo dynamics \cite{aldous} and its relation to
observables, i.e., the equilibrium and nonequilibrium relaxation
functions.  Let $W$ be a transition matrix of an irreducible,
aperiodic, and reversible Markov chain with an equilibrium (invariant)
probability distribution $p$.  We have a detailed balance equation
between $W$ and $p$,
\begin{equation}
 p_i W_{ij} = p_j W_{ji}.
\end{equation}
This equation implies that the following matrix is symmetric:
\begin{equation}
S_{ij} = p_i^{1/2} W_{ij} p_j^{-1/2},
\end{equation}
thus, the eigenvalues $\lambda_m$ of $S$ are real. 
Due to the conservation of the total probability, it can also be
shown that $| \lambda_m| \leq 1$.  
Let the eigenvectors of $S$ be $u_{im}$ for eigenvalue
$\lambda_m$, then the left and right eigenvectors of $W$ is $x_i =
p_i^{1/2} u_{im}$ and $y_i = p_i^{-1/2} u_{im}$, respectively, such
that
\begin{equation}
x W = \lambda_m x, \qquad W y = \lambda_m y.
\end{equation}
The equilibrium distribution corresponds to $\lambda_0 = 1$, $x^{(0)}
= p$, and $y^{(0)}_{i} = 1$.  The next eigenvalue $\lambda_1$ nearest
to 1 controls the rate of convergence.  We define the exponential
relaxation time $\tau$ (in units of one Monte Carlo step or attempt)
by $\lambda_1 = \exp(-1/\tau)$.

We can represent the relaxation of a general observable $Q$ in terms
of the initial distribution $p(0)$ or equilibrium distribution $p$ and the eigenspectrum of $S$ as
\begin{eqnarray}
\langle Q(t)  \rangle_{p(0)} =  \sum_{k} \lambda_{k}^t d_k c_k,\\
\langle Q(t)Q(0) \rangle_{eq} = \sum_{k} \lambda_{k}^t c_k^2,
\end{eqnarray}
where the averages are over the initial distribution and equilibrium
distribution, respectively, and
\begin{eqnarray}
c_k = \sum_{i} p_{i}^{1/2} Q_i\, u_{ik},\\
d_k = \sum_{i} p_{i}^{-1/2}  p_i(0) u_{ik}.
\end{eqnarray}
We define the normalized relaxation function $f(t)$ to be linear in
$\langle Q(t) \rangle$ or $\langle Q(t)Q(0) \rangle$ such that
$f(0)=1$ and $f(\infty) = 0$, e.g., $f(t) = (\langle Q(t) \rangle -
\langle Q(\infty) \rangle)/ (\langle Q(0) \rangle - \langle Q(\infty)
\rangle)$.  The integrated correlation time is defined as
\begin{equation}
\tau_{\rm int} = \sum_{t=0}^\infty f(t).
\end{equation}
We note that the integrated correlation time depends not only on the
observable $Q$ but also on the dynamics and the full eigen spectrum.
The integrated correlation time for the equilibrium correlation and
nonequilibrium relaxation is not the same.  On the other hand, the
exponential relaxation time, defined in the large time limit, $f(t)
\sim \exp(-t/\tau)$, is an intrinsic property of the Markov chain.  It
is the same for both the equilibrium and nonequilibrium situations.

\section{Exact Calculation in One Dimension}
It is instructive to look at the eigen spectra of the Swendsen-Wang
and Wolff dynamics in one dimension (1D).  We consider the
Swendsen-Wang dynamics in 1D with an open boundary condition.
Consider spin $\sigma_i$ at a one-dimensional lattice site $i=1$, 2,
$\cdots$, $L$, $L+1$.  We define the link variable $b_i = 1 -
\delta_{\sigma_i, \sigma_{i+1}}$. The energy of the system is
\begin{equation}
E(\sigma) = - J \sum_{i=1}^L \sigma_i \sigma_{i+1} 
          = 2J \sum_{i=1}^L b_i + {\rm const}.
\end{equation}
We introduce the bond variables $n_i = 0, 1$ representing the absence or
presence of a bond in the Swendsen-Wang dynamics, then the joint
probability distribution of both the spin and bond is proportional to
\begin{equation}
P(\sigma, n) = \prod_{i=1}^L  \bigl[ p \delta_{\sigma_i, \sigma_{i+1}} 
\delta_{n_i,1} + (1-p) \delta_{n_i,0} \bigr],
\end{equation}
where $p=1-\exp\bigl[-2J/(k_BT)\bigr]$.  The marginal distribution of
the spins is given by $\sum_{n} P(\sigma, n) = P(\sigma) = \prod_i
\exp[-2Jb_i/(k_B T)]$.  The distribution of the bonds is the special
case of the Fortuin-Kasteleyn formula, $P(n) = \sum_{\sigma} P(\sigma,
n) = p^{N_b} ( 1-p)^{L-N_b} 2^{L-N_b+1}$, $N_b = \sum_{i} n_i$.  The
Swendsen-Wang algorithm is to apply alternatively two conditional
probabilities, sample bonds given the spins $P(n|\sigma) = P(\sigma,
n)/P(\sigma)$, and sample spins given the bonds $P(\sigma|n) =
P(\sigma, n)/P(n)$.  Instead of using the spin variables, it is more
convenient to use the link variable $b_i$.  In terms of $b_i$, the
conditional probabilities are simple products:
\begin{eqnarray}
P(n | b ) = \prod_i f(b_i, n_i), 
\qquad f = \left( \begin{array}{cc} 1-p & p\\ 1 & 0\end{array} \right), \\
P(b | n) = \prod_i g(n_i, b_i), 
\qquad g = \left( \begin{array}{cc} 1/2 & 1/2\\ 1 & 0\end{array} \right).
\end{eqnarray}
The transition probability from a given link configuration to another
link configuration is
\begin{equation}
W(b \to b') = \sum_{n} P(b'|n) P(n|b) = \prod_{i} w_{b_i, b'_i},
\qquad w = \left( \begin{array}{cc} \frac{1+p}{2} & \frac{1-p}{2}\\ 1/2 & 1/2\end{array} \right).
\end{equation}
The matrix $w$ has eigenvalues 1 and $p/2$, with left eigenvectors
$v^{(1)}=(1,1-p)$ and $v^{(2)} = (1,-1)$, respectively.  We note that
the full matrix $W$ for the whole system is a direct product from the
contribution of each site.  Thus, the eigen values of $W$ are
$\lambda_m = (p/2)^{m}$ \cite{brower}, with $L!/(m!(L-m)!)$ fold
degenerate eigen vectors, $\prod_i v^{(k_i)}(b_i)$, $k_i=1$ or 2, with
$L-m$ terms of choices for $k_i=1$ and $m$ terms for $k_i=2$, where
$m=0,1,2,\cdots, L$.  The eigenvalue $\lambda_0 = 1$ corresponds to
the equilibrium state with a left eigenvector $P(\sigma)$.  The next
eigenvalue $\lambda_1=p/2 = \exp(-1/\tau)$ gives the relaxation time.
We note that rest of the decay times $\tau/m$ are well-spaced.

It is easy to write down the probability distribution of the domain
length since each link evolves independently.  Let $P_l(t)$ be the
probability for observing a domain of a length $l$ with $-$ spins
terminated by $+$ spins, then
\begin{equation}
P_l(t) = p(b=1,t) \bigl[ 1 - p(b=1,t) \bigr]^{l-1},
\end{equation}
where $p(t)= p(0)w^t$ is the probability of the link variable takes
the value $1$ at step $t$.  Similar result using a continuous time
dynamics is given in Ref.~\cite{krapivsky}.

\begin{figure}
\includegraphics[%
  scale=0.4]{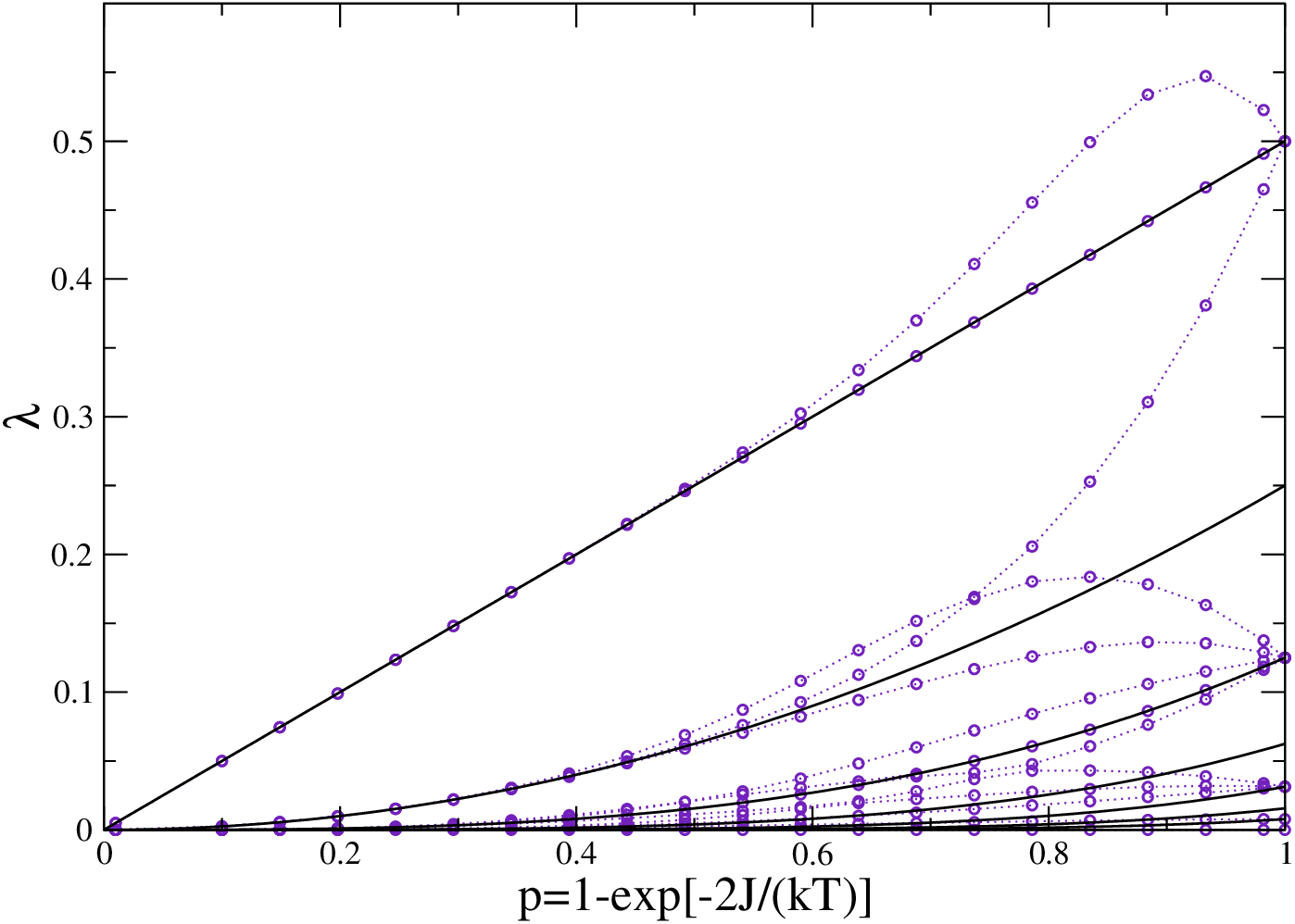}
\caption{\label{fig:spectra}Eigenvalues of the transition matrix of
the Swendsen-Wang dynamics in the 1D Ising chain with the open
boundary condition (solid lines) and the periodic boundary condition
(dotted line with circles) for a lattice size $L=8$.}
\end{figure}

When a periodic boundary condition is used, we are no longer able to
find the eigen spectrum analytically.  In Fig.~\ref{fig:spectra}, we
show the numerical results by diagonalizing the transition matrix on
an $L=8$ chain with the periodic boundary condition and compare with
the open boundary condition result.  We can make several interesting
observations.  The eigenvalue $\lambda_1 = p/2$ is always present for
both periodic and open boundary conditions for any lattice size $L$.
However, we are unable to prove it rigorously.  With the periodic
boundary condition, the eigen value $p/2$ no longer corresponds to the
slowest mode.  It seems reasonable that the simple spectrum $(p/2)^m$
is a good approximation if the correlation length $\xi \sim
\exp(2J/(k_BT))$ is much smaller than the system size $L$, thus it is
the correct spectrum in the thermodynamic limit.  For finite sizes
when $\xi$ is comparable to size $L$, the degenerate spectrum splits
and rejoins at $T=0$.
 
\section{Simulation Setup}

The two-dimensional (2D) Ising model is prepared initially in a random
state with a zero magnetization or at the ground state, but evolves at
the critical temperature. After a certain time before reaching 
equilibrium, one can observe an exponential decay of the system
energy, which satisfies the following equation:
\begin{equation}
E(t)\approx A e^{-t/\tau}+E(\infty),
\label{eq:1}
\end{equation}
where $\tau$ is the so-called exponential relaxation time. The
exponential relaxation time $\tau$ is an intrinsic property of the
Monte Carlo algorithm, which is defined by the first excited
eigenvalue $\lambda_1$ of the transition matrix, and should be
independent of the initial states in the simulations. At the
transition temperature, the dynamic scaling theory predicts that
$\tau$ diverges according to $\tau\sim L^{z_{\rm exp}}$ in the
thermodynamic limit. This defines the exponential dynamic critical
exponent $z_{\rm exp}$. On the other hand, the integrated correlation
time is defined as \cite{landau-binder}:
\begin{equation}
\tau_{{\rm int}}=\sum_{t=0}^{\infty}\frac{E(t)-E(\infty)}{E(0)-E(\infty)}.
\label{eq:2}
\end{equation}
Similarly, one may define the integrated dynamic critical exponent
$z_{\rm int}$ through $\tau_{{\rm int}}\sim L^{z_{\rm int}}$.
In order to get a more accurate result, we use the exact value of
$E(\infty)$ for the 2D Ising model \cite{fisher}. 

The Wolff algorithm exhibits an important difference compared to other
algorithms. It only updates the spins belonging to a certain cluster
around the seed spin at each Monte Carlo step, while other algorithms
sweep the whole lattice.  For a fair comparison with the
Swendsen-Wang or single-spin-flip algorithms, we need to rescale the
Wolff Monte Carlo steps.  Specifically, the Monte Carlo time $t'$ in
the Wolff algorithm should be transformed to $t$
\begin{equation}
t=\sum_{t''=1}^{t'}\frac{C(t'')}{L^d}
\label{eq:3}
\end{equation}
where $t'$ or $t''$ is the Monte Carlo time step of the Wolff single
cluster flip (i.e. the number of clusters flipped so far), $C(t'')$ is
the average size of the cluster at step $t''$, and $L^d$ ($d=2$) is
the total number of spins of the system. $t$ is proportional to the
actual CPU time.  This newly scaled time $t$ should then be used in
Eq.~(\ref{eq:1}) for the exponential relaxation time of the Wolff
dynamics.

The integrated correlation time should also be changed to
\begin{equation}
\tau_{\rm int}=\sum_{t'=0}^{\infty}\frac{E(t')-E(\infty)}{E(0)-E(\infty)}
\times \frac{C(t')}{L^d}.
\label{eq:4}
\end{equation}
In the Swendsen-Wang algorithm, there is no complication in the
definition of time, we straightforwardly use Eq.~(\ref{eq:1}) and
(\ref{eq:2}) to calculate the dynamic exponent $z_{\rm exp}$ and
$z_{\rm int}$.

Our definition of the transformed time $t$ is slightly different from
that in Ref. \cite {gunduc}. In Ref. \cite {gunduc}, the time $t$ is
defined by $t=t'C(t')$. If $C(t')$ takes a scaling form $C(t') \propto
t'^{\alpha}$, two definitions coincide. However, the disadvantage of
the definition in \cite {gunduc} is that one needs to assume or know
the scaling behavior of the cluster size $C(t')$. Our definition of
the time $t$ has clear physical meaning, i.e., whenever the number of
the flipped spins reaches $L^d$, it counts as a unit time.


\section{Results}

We use the standard Hamiltonian of the two-dimensional Ising model,
\begin{equation}
-\beta H=K\sum_{\langle i j \rangle}\sigma_{i}\sigma_{j}.
\label{eq:5}
\end{equation}
Here $\beta=1/(k_BT)$ and $K=J/(k_BT)$, $k_B$ is the Boltzmann
constant, $T$ is temperature and $J$ is the interaction energy between
two spins.  Spins $\sigma_{i}$ take only values $+1$ and $-1$.  The
site $i$ is on a square lattice with periodic boundary conditions.

\begin{figure}
\includegraphics[%
  scale=0.4]{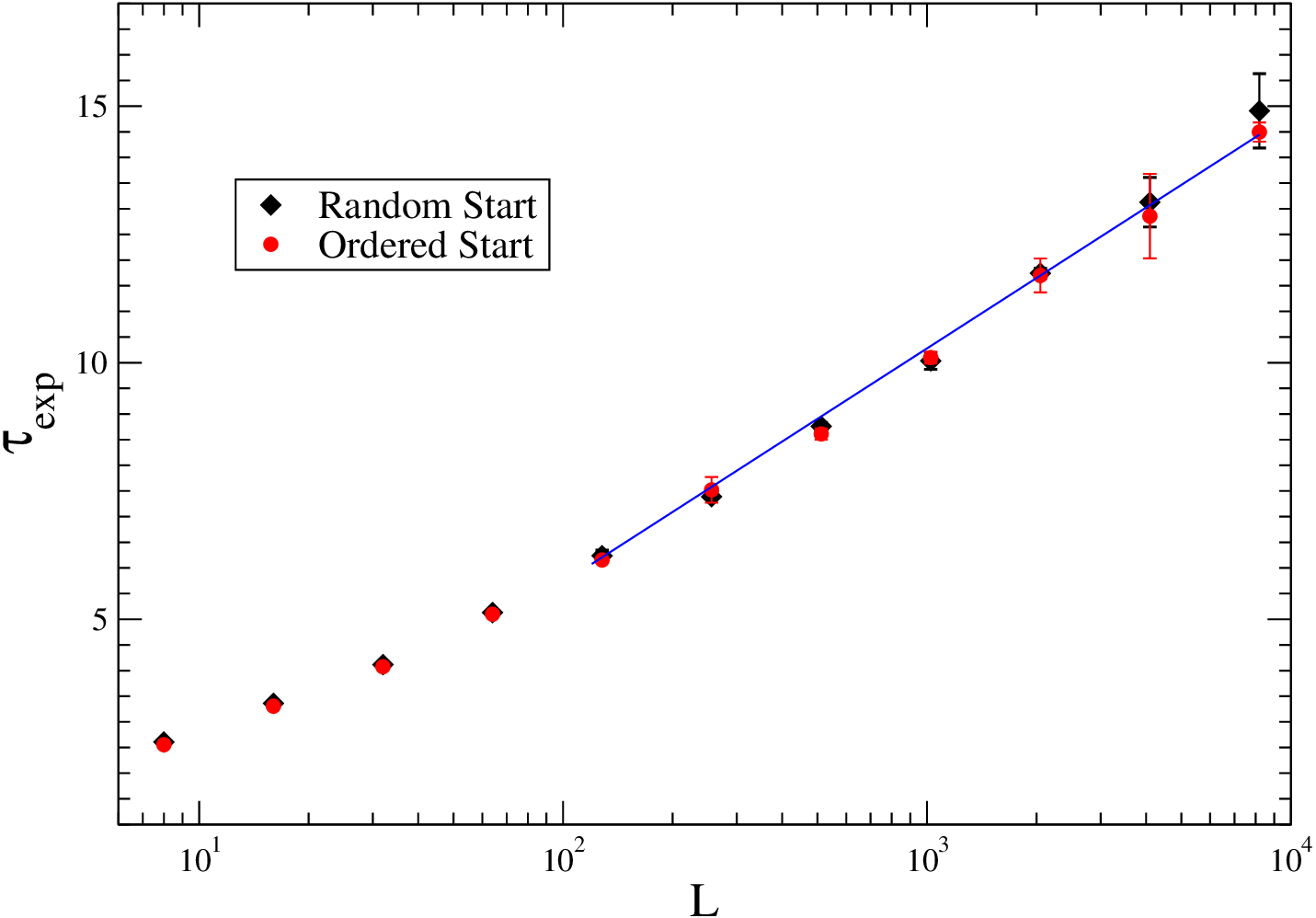}
\caption{\label{fig:swexptau}$\tau_{\rm exp}$ versus $L$ in linear-log
scale for the 2D Ising model with the Swendsen-Wang dynamics. The
diamonds are for random initial configurations, and circles are for
the ordered initial state.  A straight-line fit gives $\tau_{{\rm
exp}} = -3.42 + 1.98 \ln L$. }
\end{figure}

Let us start our numerical simulations with the Swendsen-Wang
algorithm.  To calculate the dynamic exponent $z_{\rm exp}$ of the
Swendsen-Wang algorithm, we have used lattices $L= 64$, 128, 256, 512,
1024, 2048, 4096, 8192, and each has $2^{24}$, $2^{21}$, $2^{19}$,
$2^{17}$, $2^{15}$, $5\times10^{4}$, $4\times 10^4$, $2\times10^4$
runs respectively, and the maximum Monte Carlo time steps of each
lattice are 60, 70, 70, 80, 90, 100, 100, 110.  Here two different
initial temperatures $T=\infty$ and $T=0$ are used in the
simulations. The system evolves at the critical temperature.  From the
exponential decay of the system energy in Eq. (\ref {eq:1}), one
measures the relaxation time $\tau$. The results are plotted on
a linear-log scale in Fig.~\ref{fig:swexptau}. The relaxation times
are almost identical for both the hot start and cold start. Obviously,
an approximate linear behavior is observed for the lattice size $L >
100$ in Fig.~\ref{fig:swexptau}. In other words, a logarithmic
dependence $\tau \sim \ln L$ gives a better fit to the numerical
data. If we take into account the data of smaller lattices, the
$L$-dependence of the relaxation time could be given by $\tau
\sim (\ln L)^{1.2}$.

\begin{figure}
\includegraphics[%
  scale=0.4]{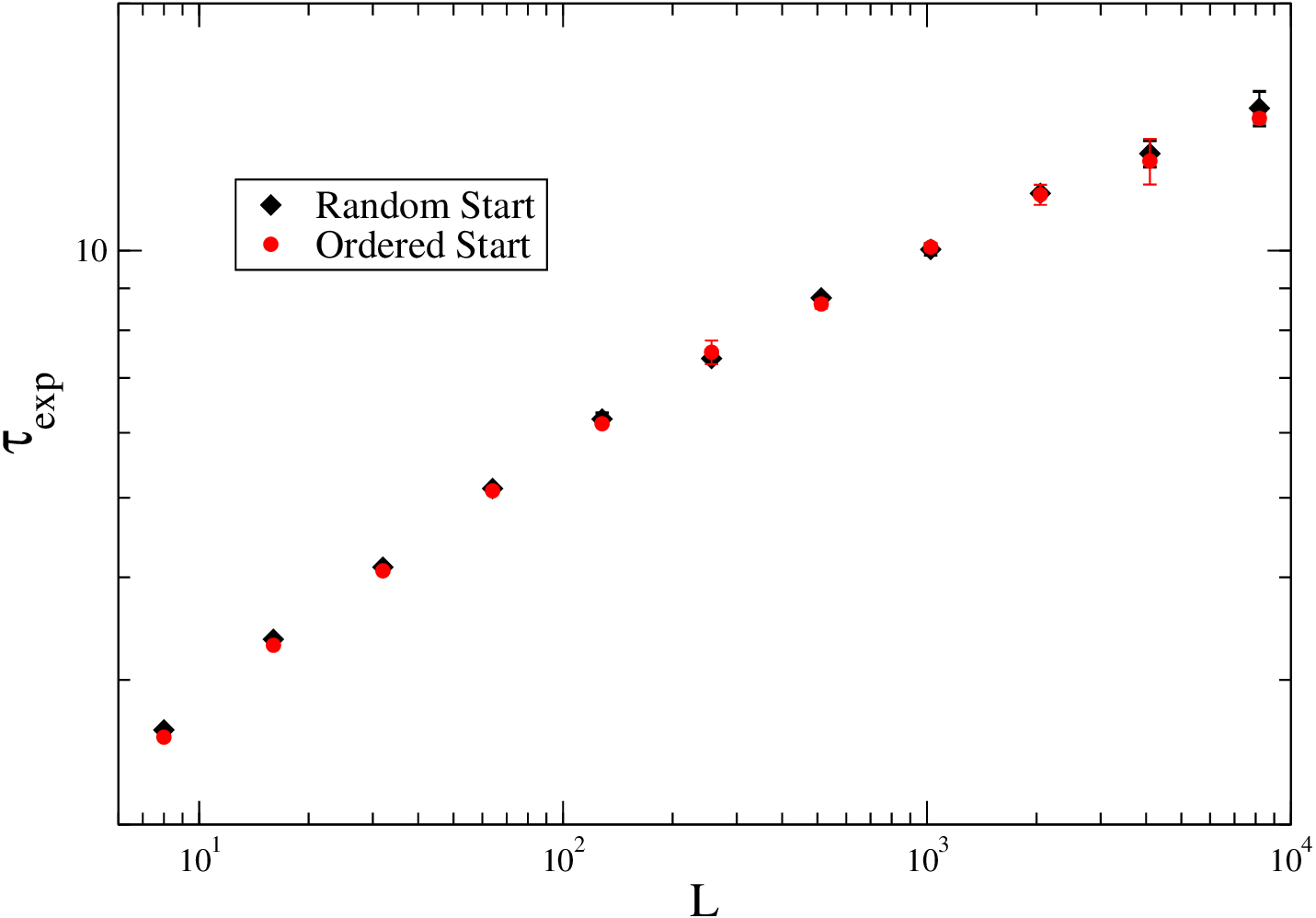}
\caption{\label{fig:swexpll}$\tau_{\rm exp}$ versus $L$ in double
logarithmic scale for the 2D Ising model with the Swendsen-Wang
dynamics.  }
\end{figure}

If we analyze the data in the form of a power-law dependence, we found
that the effective exponent $z_{\rm exp}$ decreases continuously with
increasing lattice sizes, reaching 0.18 at the largest size
simulated. This is shown with a log-log scale in
Fig.~\ref{fig:swexpll}, and it strongly suggests that the relaxation
time does not follow a power law with a small but finite exponent.
Our extensive data with large system sizes thus agree with the
conclusion of Heermann and Burkitt \cite{heermann}.  Previous
calculations \cite{sw,wolff-z,baillie,kerler,kozan,sokal} gave
various values ranging from 0.35 to 0.2.  This appears to be an effect
of finite lattice sizes.

For comparison, we also calculated the integrated relaxation time for
the Swendsen-Wang dynamics, it is nearly a constant around $\tau_{\rm
int} \approx 3.1$.  This implies that $z_{\rm int} \approx 0$, it is
consistent with $z_{\rm exp}$.

\begin{figure}
\includegraphics[%
  scale=0.4]{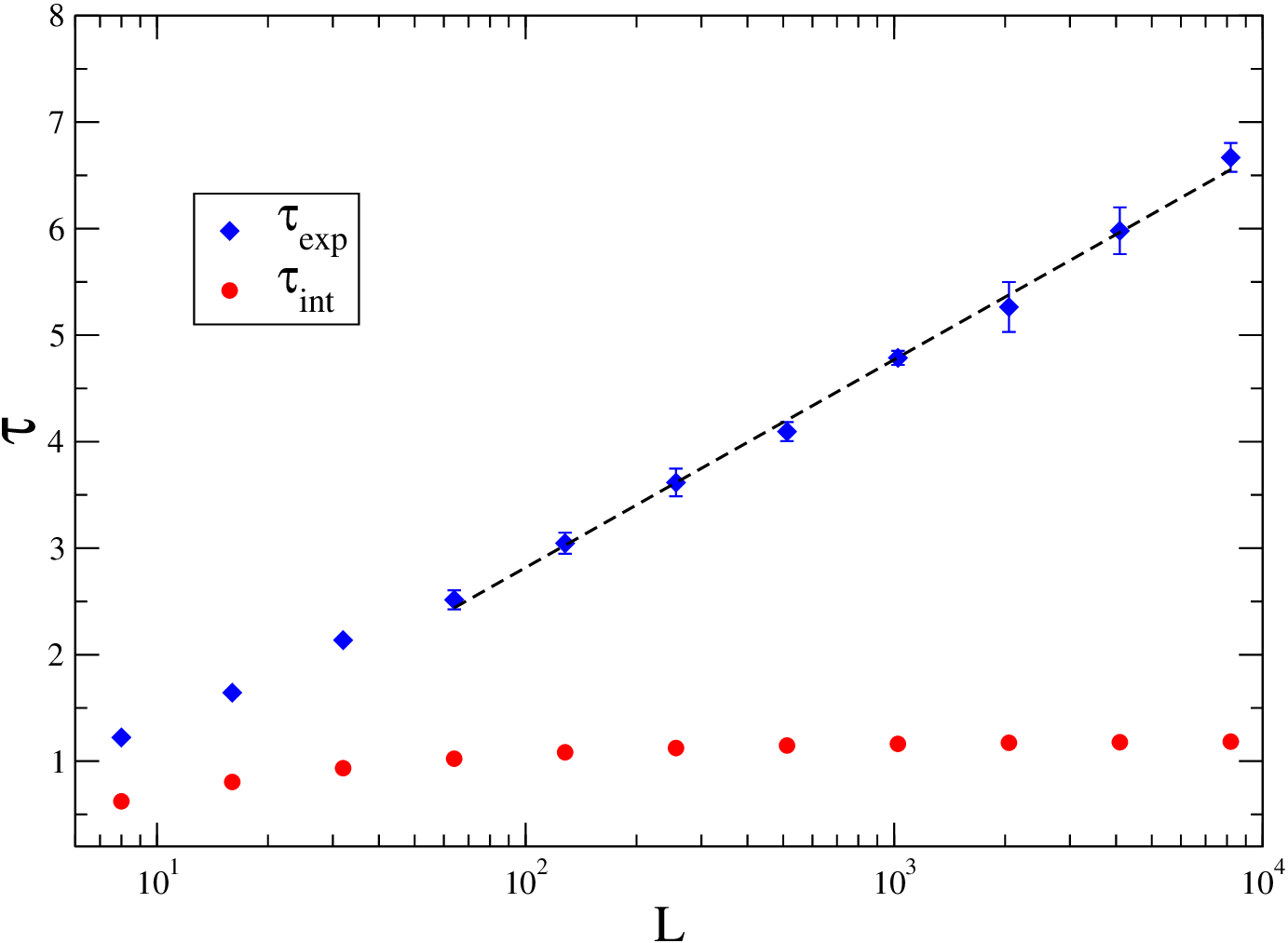}
\caption{\label{fig:LTwolff}$\tau$ versus $L$ in linear-log scale for
the 2D Ising model with the Wolff dynamics starting from an ordered
initial state.  The diamonds are exponential relaxation times, and a
straight-line fit gives $\tau_{{\rm exp}}= 1.09 + 0.85 \ln L$. The
circles are integrated relaxation times. They reach a steady value
near $1.17$ for lattice sizes larger than $256$. }
\end{figure}

We now investigate the Wolff dynamics with the fully ordered state at
$T=0$ as the initial state, and evolve the system at the critical
temperature.  The lattice sizes are $L = 8$, 16, 32, 64, 128, 256,
512, 1024, 2048, 4096, 8192 with $2^{29}$, $2^{27}$, $2^{24}$,
$2^{21}$, $2^{20}$, $2^{20}$, $2^{17}$, $5\times 10^{5}$,
$2\times10^{5}$, $8\times10^{4}$, $2\times10^{4}$ independent runs,
respectively.  The maximum Monte Carlo time steps of each lattice are
80, 100, 150,180, 200, 200, 200, 220, 250, 250, 250. We observe that
the dynamic behavior here is very similar to that of the Swendsen-Wang
dynamics.  As shown in Fig.~\ref{fig:LTwolff}, the correlation time
exhibits a logarithmic size dependence even from a relatively small
lattice size.  The integrated relaxation time is nearly a constant,
$\tau_{\rm int} \approx 1.17$ when the lattice size is larger than
$256$. This implies that $z_{\rm int} \approx 0$.  It is intuitively
understandable that the Wolff dynamics starting from a fully ordered
state is similar to the Swendsen-Wang dynamics.  When the
Swendsen-Wang algorithm is initialized in the order state ($T=0$), and
then evolves at critical temperature, it forms very large clusters
which dominate the evolution of every Monte Carlo step, while other
small clusters' effect can be neglected. Thus its behavior is much
like a Wolff algorithm initializing at the ordered state, for it also
has a very large cluster dominating the dynamic evolution.  For the
Swendsen-Wang algorithm with a disordered initial state, our
simulations show that the clusters grow rapidly, and the relaxation
time is almost the same as that with an ordered initial state.

\begin{figure}
\includegraphics[%
  scale=0.4]{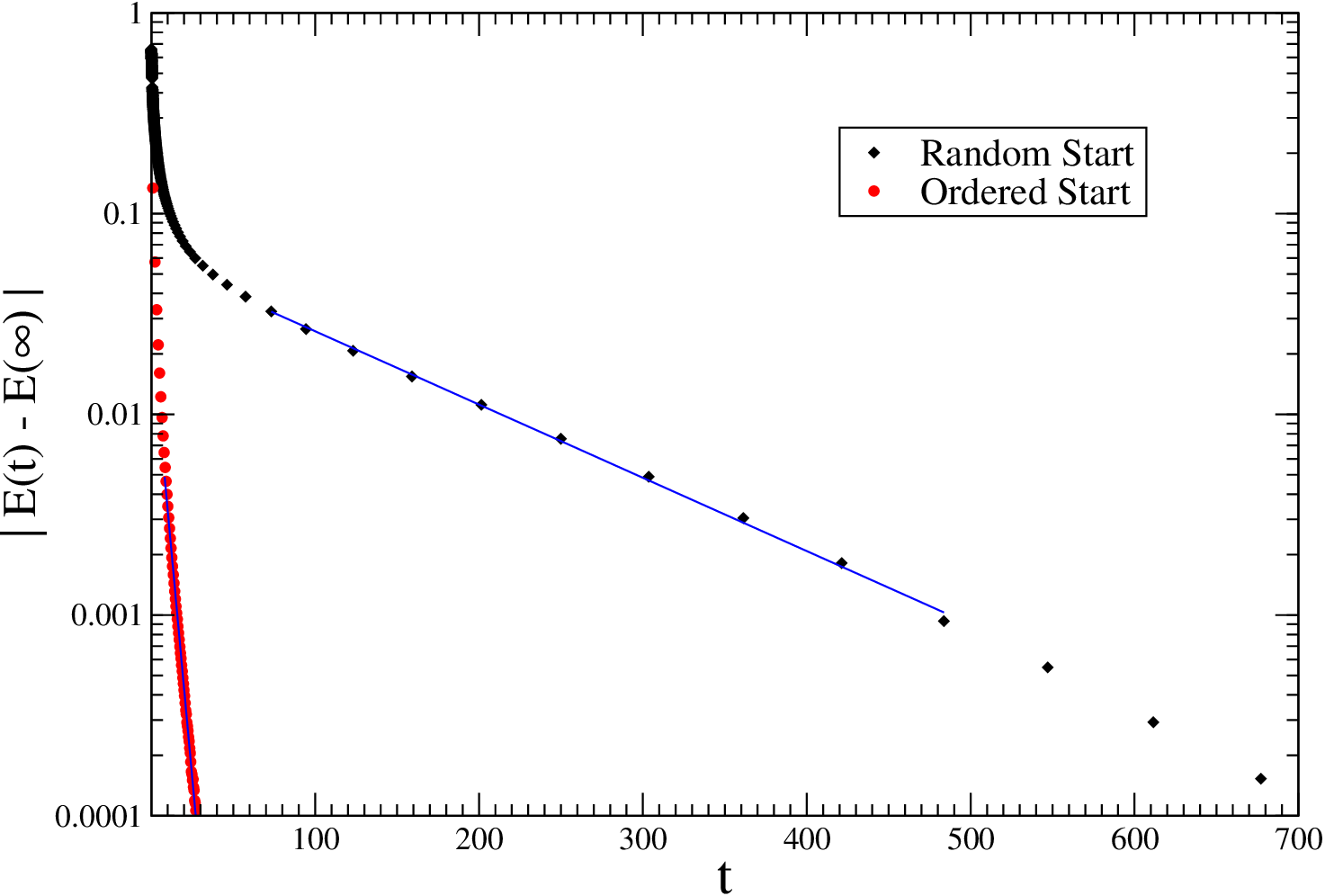}
  \caption{\label{fig:wolffdiff}$|E(t)-E(\infty)|$ versus $t$ in
   log-linear scale for the 2D Ising model with transformed $t$ for the Wolff
   dynamics. The lattice size is $L=2048$.}
\end{figure}

For Wolff dynamics with completely disordered initial states, we have used
lattices $L =64$, 128, 256, 512, 1024, 2048 with $2^{24}$, $2^{23}$,
$2^{20}$, $2^{17}$, $2^{15}$, $2^8$ runs respectively.
The maximum Monte Carlo time steps of
each lattice are 450, 1337, 4770, 18000, 69910, 300000 in the original
time unit of $t'$. 
For the Wolff algorithm with a disordered initial state, the dynamic
behavior is rather complicated. 
In order to have a better comparison of the
ordered and disordered initial starts, we present the dynamic
relaxation of the system energy for the lattice size $L=2048$ with two
different initial conditions in Fig.~\ref{fig:wolffdiff}. The figure
shows that the dynamic relaxation with a disordered start is much
slower.

\begin{figure}
\includegraphics[%
  scale=0.4]{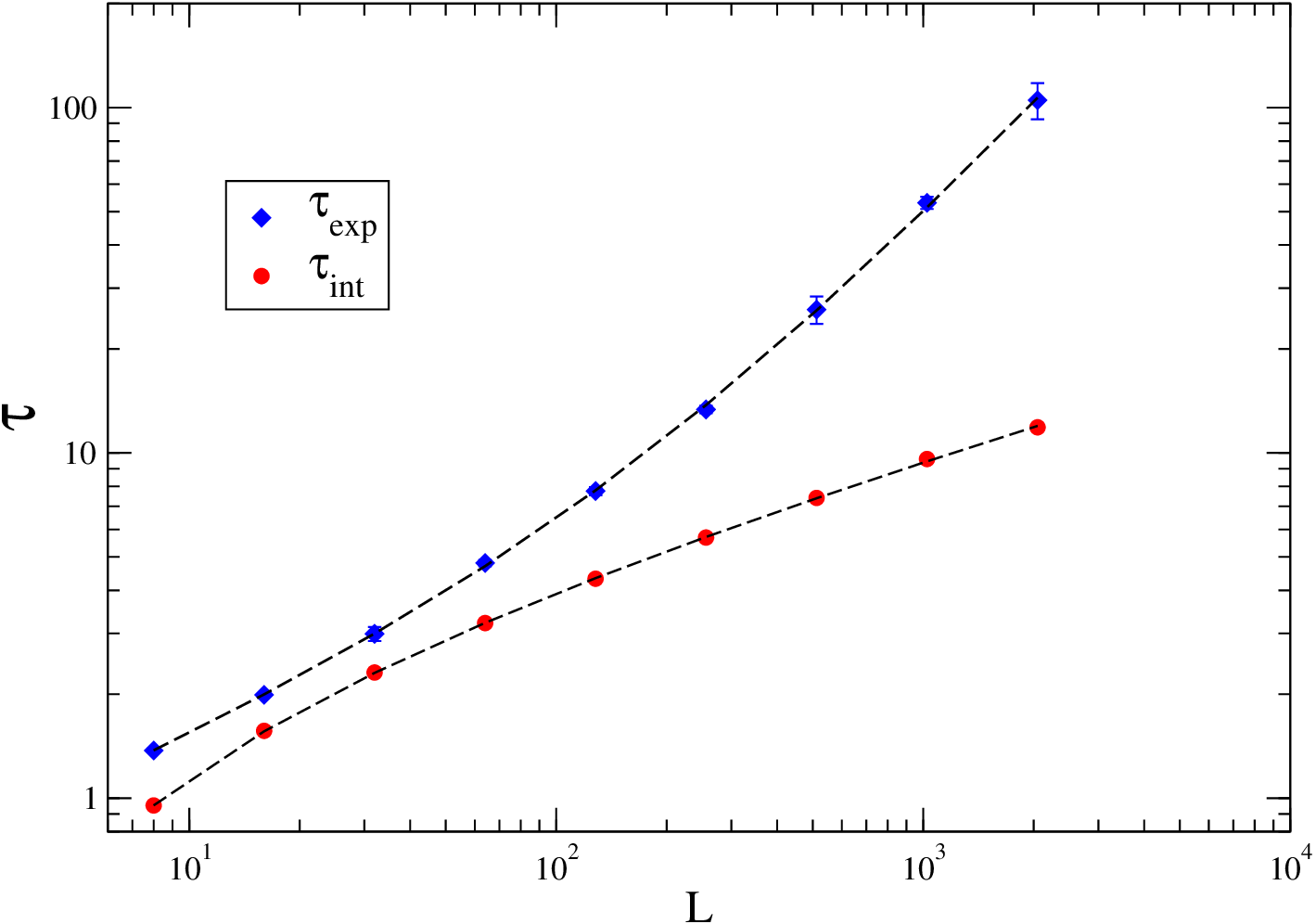}
\caption{\label{fig:wolff}$\tau$ versus $L$ in double logarithmic
  scale for the 2D Ising model with the Wolff algorithm starting from
  disordered states. The diamonds are
  exponential relaxation times $\tau_{\rm exp}$. The circles are
  integrated relaxation times.}
\end{figure}

In Fig.~\ref{fig:wolff}, $\tau_{\rm exp}$ and $\tau_{\rm int}$ are
plotted as functions of the lattice size $L$ in log-log scale. For
$\tau_{\rm exp}$, an approximate power-law behavior is observed, and
from large lattice sizes one derives a dynamic exponent $z_{\rm exp}
\approx 1.0$.  To obtain a better value of $z$, one may consider
corrections to scaling. For example, assuming $\tau_{\rm exp} \sim
L^z(1+c/L^{\delta})$, the fitted dynamic exponent is $z_{\rm exp} =
1.19(2)$. Similarly we estimate $z_{\rm int}=0.29(3)$ from $\tau_{\rm
int}$. $z_{\rm exp}$ is much larger than the results obtained in
equilibrium correlation functions \cite{wolff-z}. Both of these values
are quite different from that reported in Ref.~\cite{gunduc}, where it
is concluded that the exponent $z$ is very close to zero.

To understand the difference between our analysis and that in
Ref.~\cite{gunduc}, we have also performed the scaling plot of Fig.~1
in Ref.~\cite{gunduc} with our numerical data. Indeed, the scaling
collapse is observed for large lattices. According to our scaling
analysis, however, a dynamic exponent $z \approx 1.7$ should be
extracted. Following the definition in Ref.~\cite{gunduc}, $t=t'
\langle C(t') \rangle /L^d$ with $t'$ being the Monte Carlo time step
of the Wolff single cluster flip, and then $\tau=\tau'\langle C
\rangle/L^d$. Assuming that $\langle C\rangle$ behaves like
susceptibility, i.e., $\langle C \rangle \sim L^{\gamma/\nu}$, one may
deduce $z=z'-(d-\gamma/\nu)= z'+2(Y_H-d)$. In Ref.~\cite{gunduc}, this
is written as $z=z'-(2Y_H-d)$. Together with other inconsistent
formulations, a dynamic exponent $z \sim 0$ is derived.

In numerical simulations in equilibrium, one tends to conclude that
the dynamic exponent $z$ of the Wolff algorithm is close to zero. This
is in agreement with our results from the dynamic relaxation starting
from an ordered initial state. Compared with the dynamic relaxation of
the Wolff algorithm with an ordered initial state, why does the
dynamic relaxation with a disordered initial state show an anomalous
behavior? Our conjecture is that the eigenvalues $\lambda_m$ of the
transition matrix of the Wolff algorithm are rather dense. Within a
rather long time, the contribution of the higher eigenvalues to the
dynamic observable will not be suppressed in the dynamic relaxation
starting from a disordered state. The dynamic exponent measured in
this paper and in Ref.~\cite{gunduc} is actually an effective
one. This also explains why the dynamic exponent extracted from the
exponential decay of the system energy is somewhat smaller than that
from the dynamic scaling behavior in the relatively short time regime
in Ref.~\cite{gunduc}. Our numerical simulations and data analysis
actually show that if possible, one should avoid starting the
simulations with the Wolff algorithm from a disordered state.

Finally, we should mention that for the dynamic relaxation of the
Wolff algorithm with a disordered initial state, it is hard to measure
the relaxation time $\tau_{\rm exp}$ in an extremely long time regime
where the contribution of the higher eigenvalues is already
suppressed, since $|E(t)-E(\infty)|$ is too much fluctuating.

\section{Conclusion}

In summary, we have computed both the exponential and integrated
relaxation times for the Wolff and Swendsen-Wang algorithms, with both
disordered and ordered initial states. For the Swendsen-Wang dynamics,
the exponential relaxation time shows a logarithmic dependence on the
lattice size $L$ for both initial states, while the integrated
relaxation time tends to a constant. This is a strong evidence that
$z_{\rm exp} = 0$ for the Swendsen-Wang algorithm. For the Wolff
dynamics with an ordered initial state, the results are similar to
those of the Swendsen-Wang dynamics. For the Wolff dynamics with a
disordered initial state, however, the dynamic relaxation is very
slow, and it takes a very long time to approach equilibrium. If
one measures the relaxation time $\tau_{\rm exp}$ in a reasonable time
regime in the simulations, an effective dynamic exponent 
$z_{{\rm exp}}=1.19(2)$ is obtained.

\section*{Acknowledgements} 
This work was supported in part by NNSF (China) under Grant No. 10325520 and
by an International Collaboration Fund, Faculty of Science, NUS.  Part of
the computation was performed on the Singapore-MIT Alliance linux clusters.

\end{document}